\documentclass[showpacs,preprintnumbers,amsmath,amssymb,preprint,final]{revtex4}

\makeatletter
\def\@dotsep{4.5}
\makeatother

\newcommand{\E}{{\bf\mathrm{e}}}
\newcommand{\ud}[1]{{\,\text{d}#1}}

\begin{document} 

\preprint{PRE}

\numberwithin{equation}{section}        
\renewcommand{\theequation}{\arabic{section}.\arabic{equation}}

\title{Exhaustion of Nucleation in a Closed System}
\author{Yossi Farjoun}
 \email{yfarjoun@math.mit.edu}
\thanks{Corresponding author}
\affiliation{Department of Mathematics, MIT}
\author{John C. Neu}%
\email{neu@math.berkeley.edu}
\affiliation{Department of Mathematics, UC Berkeley}
\date{\today}

\begin{abstract}
We determine the distribution of cluster sizes that emerges from an
initial phase of homogeneous aggregation with conserved total particle
density. 
The physical ingredients behind the predictions are essentially
classical:
Super-critical nuclei are created at the Zeldovich rate, and before
the depletion of monomers is significant, the characteristic cluster
size is so large that the clusters undergo diffusion limited growth.
Mathematically, the distribution of cluster sizes satisfies an
advection PDE in ``size-space''.
During this \emph{creation} phase, clusters are nucleated and then grow to a size much larger
than the critical size, so nucleation of super-critical clusters at the
Zeldovich rate is represented by an effective boundary condition at
zero size.
The advection PDE subject to the effective boundary condition constitutes a
``creation signaling problem'' for the evolving
distribution of cluster sizes during the creation era.

Dominant balance arguments applied to the advection signaling problem
show that the characteristic time and cluster size of the creation era
are exponentially large in the initial free-energy barrier against
nucleation, $G_*$.
Specifically, the characteristic time is proportional to
$\E^{\frac25\frac{G_*}{k_BT}}$ and the characteristic number of
monomers in a cluster is proportional to
$\E^{\frac35\frac{G_*}{k_BT}}$.
The exponentially large characteristic time and cluster size give
a-posteriori validation of the mathematical signaling problem. 
In a short note, Marchenko \cite{Marchenko96} obtained these
exponentials and the numerical pre-factors, $\frac25$ and $\frac35$.
Our work adds the actual solution of the kinetic model implied by
these scalings, and the basis for connection to subsequent stages of
the aggregation process after the creation era.
\end{abstract}



\maketitle


\section{Introduction}
The traditional idea of \emph{nucleation} is the growth of clusters in
an inexhaustible monomer bath, by fluctuations over a high free-energy
barrier $G_*$.
The inexhaustibly of the monomer bath means that the
super-saturation---and hence $G_*$---are constant in time. 
In the limit $G_*/k_BT\gg1$, Zeldovich \cite{ZELD43} derived an
asymptotic result for the steady nucleation rate per unit volume,
proportional to $\E^{-\frac{G_*}{k_BT}}$.
The exponential smallness of the nucleation rate in $\frac{G_*}{k_BT}$
is the a-posteriori justification of the analysis.

The steady Zeldovich rate is only a first step towards a large picture.
Starting from pure monomer, there is a duration of so-called
``transient nucleation'' in which the nucleation rate ramps up from
zero to the steady Zeldovich value.
Asymptotic analysis of transient nucleation was carried out by
Shneidman \cite{Shneidman87} and then by Bonilla et al \cite{NBC05}.
This is the very ``beginning'', but what is the ``end'' of nucleation in a closed system? 

Qualitatively, it is clear: The super-saturation decreases due to
the growth of clusters, thereby increasing the free-energy
barrier, and eventually the nucleation of new clusters is shut down.
We refer to the duration between the aforementioned transient
nucleation and this ``shutting down'' as the ``creation era.''
We use the word ``creation'' since conventionally ``nucleation''
refers to the steady state process in an inexhaustible bath as first
studied by Zeldovich.

This paper proposes a quantitative theory of the creation era.
It is \emph{quasi-static}: We assume that the Zeldovich rate
applies even though the super-saturation slowly decreases in time, and
the free energy barrier slowly increases. 
The quasi-static assumption is affirmed a-posteriori using Shneidman's criterion \cite{shneidman85} in appendix \ref{app:qss}.
Since the nucleation rate is the exponential of a large negative
quantity $-\frac{G_*}{k_BT}$, a small relative increase in $G_*$
is sufficient to reduce the nucleation rate to a small fraction of its
initial value.
This increase in $G_*$, although small, nevertheless requires an
exponentially long time (in the initial value of $\frac{G_*}{k_BT}$)
due to the exponentially small nucleation rate.
In this exponentially long time, the largest clusters grow to
an exponentially large size, so large that their growth is diffusion
limited.

In diffusion limited growth, the number $n$ of monomers in a cluster
grows at a rate proportional to linear size, so $\dot n$ is
proportional to $n^{\frac13}$.
From the above physical framework, creation era scaling units $[t]$
and $[n]$ of time and cluster-size are derived:
\begin{equation}
[t]\propto \E^{\frac25\frac{G_*}{k_BT}},\qquad [n]\propto
\E^{\frac35\frac{G_*}{k_BT}}.
\label{eq:scales:intro}
\end{equation}
In particular $[n]\propto[t]^{\frac32}$.
This is connected to diffusion limited growth: The ODE $\dot n\propto
n^{\frac13}$ has a solution proportional to $t^{\frac32}$.
It is perhaps natural that $[t]$ and $[n]$ are exponentially large in
$\frac{G_*}{k_BT}$, but what is distinctive of the physical model are
the pre-factors $\frac25$ and $\frac35$ in the exponents in
\eqref{eq:scales:intro}.
They have in fact been seen before: In a short note, Marchenko
\cite{Marchenko96} derived the \emph{same} characteristic time and
cluster-size, with the signature $\frac25$ and $\frac35$.

Once physical scaling units are established, an asymptotic theory for
the evaluation of the cluster size distribution during the creation
era readily follows.
The mathematical form of this theory is the creation signaling
problem briefly described in the abstract. Its solution is
straightforward.
The main result is a determination of the nucleation rate as a
function of time in scale-free form.

This paper is describes the beginning of the aggregation process as a
whole.
The long-time limit of the creation era provides effective initial
conditions for successive stages of the aggregation process,
eventually making contact with the classic theory of coarsening (aka
ripening) due to Lifshitz and Slyozov. This subsequent work has been
carried out, and is being prepared for publication. 

%
\section{The physical model of Nucleation and Growth}
\label{sec:nucleation_and_growth}
\noindent The physical model used here is based on several assumptions:
\begin{enumerate}[{A}1.]
\item The initial super-saturation, and monomer chemical-potential are
  small (and positive).
\item Clusters nucleate at the Zeldovich rate, which adjusts to new
  values of super-saturation immediately.
\item The total monomer density (including monomers that form
  clusters) is conserved.
\item The cluster growth is diffusion limited.
\item Initially, there are no clusters.
\item The temperature, $T$, is fixed.
\end{enumerate}
While the growth of clusters is not a deterministic process, the
fluctuations are small in the size- and time-scales that we
 investigate here; 
They can be safely ignored throughout the creation era.
The evolution of the cluster-size distribution is modeled as an
advection PDE, with no diffusion term, in the space of time $t$, and
cluster-size $n$.  
The advection velocity is the cluster growth rate, and the nucleation
rate serves as the boundary condition at size $n=0$. 

\subsection{Super-saturation and chemical potential}
Different aspects of the physical model are related via their
dependence on the chemical-potential and super-saturation. 
Let $f_1$ denote the density of monomers in the monomer-bath, and
$f_s$ denote the saturation density, that is, the density of monomer
 that would be in equilibrium with an infinite cluster. 
The super-saturation $\phi$ and monomer chemical-potential (in units of
$k_BT$), $\eta$ are defined as
\begin{equation}
\phi = \frac{f_1-f_s}{f_s},\qquad \eta = \log \frac{f_1}{f_s}.
\label{eq:eta:phi:def}
\end{equation}
Thus, $\eta$ and $\phi$ are related by: $\phi=\E^\eta-1$. 
For  $\eta\ll1$, we have $\eta\sim\phi$. 
In this section we introduce the model using $\eta$ or $\phi$ as
appropriate, but in the rest of the paper we use $\eta$
exclusively.

\subsection{Nucleation rate}
According to  Zeldovich \cite{ZELD43}, super-critical clusters are nucleated at a rate per unit volume $j$, given by
\begin{equation}
j=\omega f_s \sqrt{\frac{\sig}{6\pi}}\E^{-\frac{\sig^{3}}{2\eta^{2}}}. \label{eq:zeldovich:rate}
\end{equation}
Here, $f_s$ is the saturation density of the
monomer, $\omega$ is the evaporation rate constant so that $\omega
n^{2/3}$ is the rate at which monomers on the surface of an
$n-$cluster leave it, and $\sig$ is the surface-tension
constant, so that $k_BT\sig n^{2/3}$ is the free-energy associated
with the surface of an $n-$cluster. 
The exponent $\frac{\sig^{3}}{2\eta^{2}}$ is the asymptotic
approximation of the free-energy barrier, $\frac{G_*}{k_BT}$.

\subsection{Growth rate}
As is shown further in this paper, the characteristic size of a cluster
during the creation era is exponentially large (in $G_*$). 
This explains our \emph{assumption} that it is large enough that the
clusters' growth is limited by the diffusion of monomers, rather than by
surface reactions\footnote{The growth-rate derived from surface
  reactions is due to Becker and D\"oring \cite{BD35}.}.
This is the basis of the Lifshitz-Slyozov formula\cite{LS61}
\begin{equation}
\dot n = d (\eta n^{\third}-\sig),\qquad d\equiv \brk{3(4\pi)^2}^{\frac13} (Dv^{\frac13}f_s)\label{eq:LS:growth}
\end{equation}
which describes the growth-rate of an $n-$cluster. 
In \eqref{eq:LS:growth}, $D$ is the diffusion constant of the monomers
and $v$ is the volume per particle in the clusters.
During the creation era, the first term in \eqref{eq:LS:growth} 
dominates the second, which is henceforth ignored. 
However, a balance between the two terms reveals the critical size,
$n_*=\brk{\frac{\sig}{\eta}}^3$, which corresponds to the maximum of
the free-energy.
Ignoring the second term in the formula for $\dot n$ \eqref{eq:LS:growth}, is equivalent to the assumption that the characteristic cluster size, $[n]$, is much larger than the critical size $n_*$.

\subsection{Advection signaling problem}

We approximate the discrete cluster-size densities with a continuous
density function $r(n,t)$. 
For small $\delta n$, the density of clusters of size between
$n$ and $n+\delta n$ is $ r(n,t)\,\delta n$. 
This approximation allows us to write an advection PDE for the cluster-size
distribution using the growth rate as the advection ``velocity''.
Mathematically, this means that the distribution $r(n,t)$ satisfies the PDE
\begin{equation}
\partial_t r +d\eta \partial_n(n^\third r) =0 \text{ in } n>0,
\label{eq:advection:PDE}
\end{equation}
where $d$ is defined in \eqref{eq:LS:growth}.
The initial conditions we assume are pure monomer, corresponding to
$r(n,0)=0$ for all $n>0$.
The flux of clusters is evidently $d\eta n^\third r$, and it must tend
to the nucleation rate as $n\goto0^{+}$, giving the effective
boundary condition:
\begin{equation}
d\eta n^\third r \goto \Omega \E^{-\frac{\sig^3}{2\eta^2}}, \text{ as
} n\goto0^{+}.
\label{eq:advection:BC}
\end{equation}
Here, $\Omega \equiv f_s\omega\sqrt{\frac{\sig}{6\pi}}$, the prefactor in the Zeldovich formula \eqref{eq:zeldovich:rate}.
The superscript $+$ indicates that the limit is taken from above.
Readers who are concerned about our use of $0$ instead of $n_*$ here
or in the next subsection, are
referred to Appendix \ref{app:n_*} for a brief discussion.

The advection PDE \eqref{eq:advection:PDE} subject to the effective boundary conditions \eqref{eq:advection:BC} and the initial condition $r(n,\,0)=0$ constitute the creation signaling problem mentioned in the abstract.
It determines the evolution of $r(n,\,t)$ as a functional of $\eta=\eta(t)$.
The model is ``closed'' by a determination of $\eta(t)$ as a functional of $r(n,\,t)$ using the conservation assumption.

\subsection{Conservation of total monomer density}

The conservation of monomers is expressed (approximately) by
\begin{equation}
  f = f_1 + \int\limits_{0}^\infty n\,r(n,t)\ud{n}. 
  \label{eq:continuum:conserv}
\end{equation}
Here, the total monomer density $f$, a constant, is the sum of monomer density, $f_{1}$ in the bath, and $\int_{0}^{\infty}nr(n,\,t)\ud{n}$, which approximates the density of monomers in clusters.
Inserting the relation $f_{1}\sim(1+\eta)f_{s}$, which follows from \eqref{eq:eta:phi:def} in the limit $\eta\ll1$, into \eqref{eq:continuum:conserv}  we find  
\begin{equation}
  f = (1+\eta)f_s + \int\limits_{0}^\infty n\,r(n,t)\ud{n}.
\end{equation}
For a full derivation and further discussions on these models, we refer the readers to Wu's review article \cite{Wu96} and references therein.
\section{Asymptotic solution of the creation signaling problem}
\label{sec:creation:sol}
The equations describing the creation era advection signaling problem are gathered together
for quick reference:
\begin{align}
\partial_t r &+\partial_n(d\eta n^\third r) =0 \text{ in } n>0,
\label{eq:advection}\\
r(n,0)&=0, \text{ for } n>0, \label{eq:advection:IC}\\
d\eta n^\third r &\goto j = \Omega \E^{-\frac{\sig^3}{2\eta^2}}
\text{ as } n\goto 0^{+},\label{eq:nucleation}\\
\eta(t) &= \eta(0) - \oneover{f_s} \int\limits_{0}^\infty n r(n, t)
\ud{n}.\label{eq:conservation}
\end{align}
We assume that the initial super-saturation is small, and take
$\eps \equiv\eta(0)\ll1$
as the small parameter of the asymptotics.
\subsection{Dominant balance scalings}
In the limit \mbox{$0<\eta\ll1$}, the nucleation rate in \eqref{eq:nucleation} is
highly sensitive to small changes in the chemical potential
\mbox{$\eta$}.
Hence, we work with the \emph{change} in chemical potential, 
\begin{equation*}
\delta\eta \equiv \eta-\eta(0) = \eta-\eps,
\end{equation*}
and expect the super-saturation \mbox{$\eta$} to remain close to its original
value \mbox{$\eta(0)$} throughout the creation era.
We find scaling units \mbox{$\Brk{t}, \Brk{n}, \Brk{r}$}, and
\mbox{$\Brk{\delta\eta}$} of the variables \mbox{$t,\,n,\,r,$} and \mbox{$\delta\eta$} from
\emph{dominant balances} applied to equations
(\ref{eq:advection}--\ref{eq:conservation}).

The dominant balance associated with the advection PDE
\eqref{eq:advection} is 
\begin{equation}
\oneover{\Brk{t}}=\oneover{\Brk{n}} (d\eps\Brk{n}^\third).\label{eq:advection:DB}
\end{equation}
Integrating \eqref{eq:advection} from \mbox{$n=0$} to
\mbox{$\infty$} and using boundary condition \eqref{eq:nucleation}
we find
\mbox{$\deriv{}{t} \int\limits_{0}^\infty r\ud{n} = \Omega \E^{-\frac{\sig^3}{2\eta^2}} $},
with corresponding dominant balance 
\begin{equation}
\frac{\Brk{r}\Brk{n}}{\Brk{t}} = \Omega \E^{-\frac{\sig^3}{2\eps^2}}. \label{eq:nucleation:DB}
\end{equation}
The conservation equation \eqref{eq:conservation} can be written as
\mbox{$\delta\eta = \frac{-1}{f_s} \int\limits_{0}^\infty n\, r(n, t) \ud{n}$},
with corresponding dominant balance 
\begin{equation}
\Brk{\delta\eta} = \oneover{f_s} \Brk{r}\Brk{n}^2.\label{eq:conservation:DB}
\end{equation}
In addition to the three ``straightforward'' dominant balances
(\ref{eq:advection:DB}--\ref{eq:conservation:DB}) which
follow directly from (\ref{eq:advection}--\ref{eq:conservation}),
there is one that quantifies the change \mbox{$\delta\eta$} in chemical
potential required to  ``shut off'' nucleation.
For \mbox{$\delta\eta\ll\eta$}, the relative change in nucleation rate
\mbox{$j$} in \eqref{eq:nucleation} that results from a change \mbox{$\delta\eta$} in \mbox{$\eta$} is
\mbox{$\frac{\delta j}{j} \sim \frac{\sig^3}{\eps^3}\delta\eta,$}
so the scaling unit \mbox{$\Brk{\delta\eta}$} of \mbox{$\delta\eta$} that
corresponds to a significant change in the nucleation rate is 
\begin{equation}
  \Brk{\delta\eta} = \frac{\eps^3}{\sig^3}.\label{eq:delta:eta:DB}
\end{equation}
This is consistent with our expectation that it is small relative to $\eta$. 
We substitute \eqref{eq:delta:eta:DB} into \eqref{eq:conservation:DB}, and
then solve (\ref{eq:advection:DB}--\ref{eq:conservation:DB}) for the scaling units \mbox{$\Brk{t}, \Brk{n},$}
and \mbox{$\Brk{r}$} of time, cluster size, and cluster-size density:
\begin{align*}
  [t]&=\brk{\frac{\eps}{\sig^2d}}^{\frac35}\brk{\frac{f_s}{\Omega}}^{\frac25}\E^{\frac25\frac{G_*}{k_BT}},\\
  [n]&=\brk{\frac{d\eps^4 f_s}{\sig^3\Omega}}^{\frac35}\E^{\frac35\frac{G_*}{k_BT}},\\
  [r]&=\brk{\frac{\Omega^2\sig}{d^2\eps^3}}^{\frac35}f_s^{\frac15}\E^{-\frac65\frac{G_*}{k_BT}}.
\end{align*}
Recalling the definitions of $d$ and $\Omega$, these become
\begin{empheq}[innerbox=\fbox]{align}
\Brk{t}&=(8\pi)^{-\oneover{5}}\BRK{\eps^\frac35\sig^{-\frac75}}\E^{\frac25\frac{G_*}{k_BT}}\brk{D^3vf_s^3\omega^2}^{-\frac15},
  \label{eq:scale:t}\\
\Brk{n}&=(\pi^\frac{7}{10}2^\frac{11}{10}\sqrt{3})\BRK{\frac{D \eps^4 f_s v^{\frac13}}{\sig^{\frac{7}{2}}\omega}}^{\frac35}\E^{\frac35\frac{G_*}{k_BT}},\label{eq:scale:n}\\
\Brk{r}&=(3\cdot 2^{11}\pi^7)^{-1/5}\BRK{\frac{\sig^2\omega^2 }{\eps^3 D^2f_s^2v^{\frac23}}}^{\frac35}\E^{-\frac65\frac{G_*}{k_BT}}(f_s)\label{eq:scale:r}.
\end{empheq}
In the exponents, \mbox{$G_*= k_BT \frac{\sig^3}{2\eps^2}$} is the initial free energy
barrier.
\subsection{The Reduced Kinetics}
We nondimensionalize the creation signaling problem 
(\ref{eq:advection}--\ref{eq:conservation}) using the units in (\ref{eq:delta:eta:DB}--\ref{eq:scale:r}). 
In the limit \mbox{$\eps\goto0$}, the reduced equations are:
\begin{align}
   \partial_tr&+\partial_n(n^\third r)=0,\quad \text{ in } n>0,
\label{eq:scaled:pde}\\
    n^{\third}r&\goto  \E^{\delta \eta},\quad \text{ as } n\goto0^+,\label{eq:BC1}\\
    \delta\eta &= -\int\limits_0^\infty n\, r\ud{n}. \label{eq:nucleation:conservation}
\intertext{The initial condition is}
    r(n,0) &=0.\label{eq:IC:r}
\end{align}
The reduced signaling problem  (\ref{eq:scaled:pde}--\ref{eq:IC:r}) is
transformed into an integral equation for \mbox{$\delta\eta$}, which is
solved numerically.
The density \mbox{$r(n, t)$} is subsequently recovered from \mbox{$\delta\eta(t)$}.

\begin{figure}[ht]
\begin{center}
\centerline{
\resizebox{6cm}{!}{\includegraphics{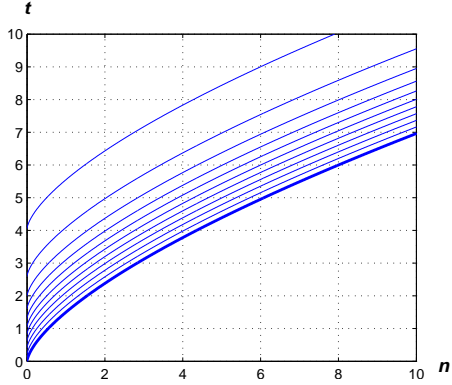}}}
\caption[Characteristic during the nucleation era]{The characteristics of the reduced PDE\@. The flux \mbox{$n^{\third}r$} is
 constant along the characteristics. The density of the curves is
 proportional to the solution \mbox{$r(n,\,t)$}.}\label{fig:char}
\end{center}
\end{figure}
The flux of super-critical clusters, \mbox{$n^\third r$},
is constant along the characteristics
\begin{equation}
  C_\tau \equiv\BRK{ \brk{\brk{\frac23(t-\tau)}^\frac32,\,t}:\quad t\ge\tau}, \label{eq:world:line}
\end{equation}
of the PDE \eqref{eq:scaled:pde}.
The characteristics $C_\tau$ can be seen in Fig.~\ref{fig:char},
in which the (horizontal) density of the characteristics
at each point \mbox{$(n, t)$} is proportional to the density of clusters of
size \mbox{$n$} at time \mbox{$t$}.
The curves in  \eqref{eq:world:line} describe  the
world-lines of clusters that nucleate at time \mbox{$t=\tau$}.
The region below the thick line in Fig.~\ref{fig:char} corresponds
to \mbox{$t<\frac32n^\frac32$}.
In this region, there are no cluster world lines and \mbox{$r(n,\,t)=0$}.

For a \emph{known} \mbox{$\delta\eta(t)$}, the solution  $r(n,\,t)$ that
has a constant flux \mbox{$n^\third r$} along characteristics, and
satisfies the BC \eqref{eq:BC1}, is
\begin{equation}
  r(n,\,t)= \left\{
\begin{aligned}
&n^{-\third}\E^{\delta\eta\brk{t-\frac32n^{2/3}}}, &t \ge \frac32n^\frac23,\\
&0, & 0\le t <\frac32n^\frac23.
\end{aligned}
\right.
 \label{eq:char:sol}
\end{equation}
An integral equation for \mbox{$\delta\eta(t)$} is found by substituting \eqref{eq:char:sol} for \mbox{$r(n,\,t)$} in the conservation identity \eqref{eq:nucleation:conservation}: 
\begin{equation}
\boxed{\delta\eta(t) = -\int\limits_0^t\brk{\frac23(t-\tau)}^\frac32\E^{\delta\eta(\tau)}\ud{\tau}.}\label{eq:integral_1}
\end{equation}
In \eqref{eq:integral_1} the variable of integration has been changed from \mbox{$n$} to \mbox{$\tau\equiv t-\frac32n^{2/3}$}.
\subsection{Creation Transition and Physical Predictions}

\begin{figure}[ht!]
\begin{center}
\resizebox{5.5cm}{!}{\includegraphics{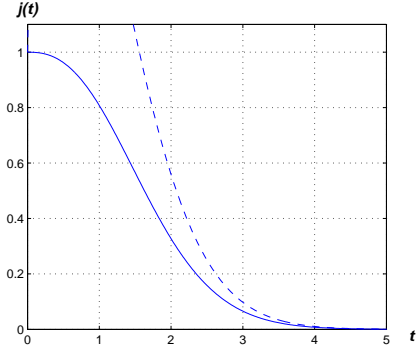}}
\caption[Nucleation rate as a function of time]{The Zeldovich nucleation
  rate \mbox{$j$} as a function of time. After an \mbox{$\O(1)$} time,
  the super-saturation
  decreases slightly, and the nucleation rate gets turned off.}
\label{fig:flux}
\end{center}
\end{figure}

\begin{figure}[ht!]
\begin{center}
  \resizebox{5.5cm}{!}{\includegraphics{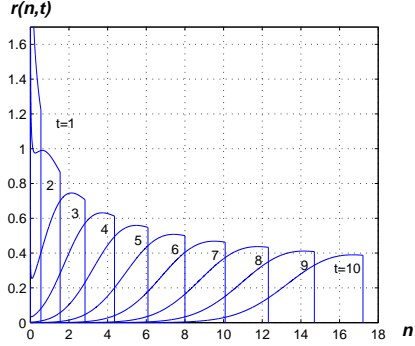}}
\caption{The density of cluster sizes, \mbox{$r(n,\,t)$}, for
  various values of \mbox{$t$}.}
\label{fig:movie}
\end{center}
\end{figure}

We solve (\ref{eq:integral_1}) numerically.
A short discussion of the method and numerical result can be found
in Appendix \ref{sec:numerical:explain}.
The nucleation rate \mbox{$j=\E^{\delta\eta(t)}$} is calculated and plotted in
Fig.~\ref{fig:flux}.
At time \mbox{$t=2$}, $j$ is about a third its original value,
and at \mbox{$t=5$} it has effectively vanished.
The distribution \mbox{$r(n, t)$} of cluster sizes is recovered from
\mbox{$\delta\eta(t)$} via  \eqref{eq:char:sol}.
Figure~\ref{fig:movie} shows \mbox{$r$} vs.\ \mbox{$n$} for an increasing sequence
of \mbox{$t$}.


From the  numerical solution  for \mbox{$j$} we find  the total
density of clusters generated during the creation era.
This density is given by the integral
$ \int\limits_0^\infty r\ud{n}$.  
Using the same change of variables from \mbox{$n$} to \mbox{$\tau$} as in \eqref{eq:char:sol} and PDE \eqref{eq:scaled:pde}, we convert 
 the integral of \mbox{$r$} into an integral of \mbox{$j=\E^{\delta\eta}$}, 
\begin{equation}
  R\equiv\int\limits_0^\infty r\ud{n} = \int\limits_0^\infty j(\tau)\ud{\tau}.\label{eq:def:R}
\end{equation}
The value of \mbox{$R$}, based on the numerical approximation to \mbox{$j(t)$}, is
\begin{equation*}
  R\Approx 1.7109.
\end{equation*}
Converting back to original physical units, the total density of
clusters produced during the creation era is
\begin{equation}
  \boxed{R[r][n]=\frac{3^{\frac{3}{10}}R}{2^{\frac{11}{10}}\pi^{\frac{7}{10}}}
  \BRK{\frac{\eps \omega}{ f_s D v^\frac13 \sig^{\frac32}}}^{\frac35} \E^{-\frac35
    \frac{G_*}{k_BT}} (f_s).} \label{eq:scales:density}
\end{equation}
\subsection{The Long-Term Behavior}
We discuss the long-term behavior of the cluster-size distribution, $r(n,\,t)$,  is determined by the long-term behavior of $j$ through Eq.~\eqref{eq:char:sol}. 
For $t$ large relative to the ``main support'' of $j=\E^{\delta\eta}$, we approximate the integral equation \eqref{eq:integral_1} by 
\begin{equation}
-\delta\eta\sim\brk{{\textstyle \frac23t}}^{\frac32}R + \brk{{\textstyle \frac23t}}^{\frac12}\int_{0}^{\infty}\tau \E^{\delta\eta(\tau)}\ud{\tau}.
\end{equation}
Fig.~\ref{fig:flux} shows the graph of $j(t)$ together with this approximation. 
The second moment of $j(\tau)$ is found to be
\begin{equation}
\int\limits_{0}^{\infty}\tau\,j(\tau)\ud{\tau}\approx 1.7773.
\end{equation}

Since $j(t)$ diminishes to a small fraction of its original value after $t=5$, the main support of $r(\cdot,\,t)$ is located in the range of $n$ where $0<t-\frac32n^{\frac23}<5$. 
This can be seen most clearly from  \eqref{eq:char:sol}.
The upper end of this interval is sharp: 
At time $t$ there are no cluster sizes greater than $(\frac23 t)^{\frac32}$, but the lower end is ``soft'' since although $j(t)$ is small at $t=5$, it is not exactly zero for any $t>0$, however the super-exponential decay of $j(t)$ allows us to treat the support of $j$ (and therefore of $r$) as being well-defined despite this softness.
By comparing the width of the distribution to the size of the largest cluster $N$,
\begin{equation}
N(t)=\brk{\frac23 t}^{\frac32},
\label{eq:N:def}
\end{equation}
we show that outside of a (relatively) narrow region the distribution is exponentially small. 
In this sense we can say that the support of $r$ is concentrated in a narrow interval, and the distribution is \emph{asymptotically} monodisperse.
Describing this narrow region is our final task.

We define the distance between $n$ and the size of the largest cluster $N$: 
\begin{equation}
\delta n = N-n.
\label{eq:delta_n:def}
\end{equation}
The support of $r$ is concentrated at values of $\delta n$ for which $0<t-\frac32(N-\delta n)^{\frac23}<5$. 
Thus,  to first order in $\frac{\delta n}{N}$, the support of $r$ is concentrated where $0<\frac{\delta n}{N^{\third}}<5$. 
In other words, the width of the distribution grows like $t^{\half}$ but the size of the largest cluster (i.e., the location of the distribution) grows like $t^{\frac32}$, so the \emph{relative} width of the distribution is shrinking. 
The \emph{long-term} description of the cluster-size density is given by
\begin{equation}
r(n,\,t)= \left\{
\begin{aligned}
&N^{-\third}j\brk{\frac{\delta n}{N^{1/3}}}, &0< \frac{\delta n}{N^{1/3}}<5\\
&0, & \text{otherwise.}
\end{aligned}
\right.
\end{equation}
Here,   $N$ and $\delta n$ are defined in \eqref{eq:N:def} and \eqref{eq:delta_n:def}.
The constant 5 is used to describe the (soft) upper bound of the support of $j(t)$. 
\section{Conclusions}
\label{sec:conclusions}
The exponential dependences of the characteristic time and cluster
size upon the initial free energy barrier are based on order of
magnitude balances applied to simple, essentially classical kinetics.
They should be the most robust results of this paper.  

We mention a biophysics application: Hydrophobic proteins have been
``crystallized'' into periodic arrays for x-ray analysis.
They are implanted into a space filling cubic phase of lipid bilayer,
which acts as the ``solvent''.
Nucleation of ``protein crystals'' is observed.
The size of the cubic phase unit cell was manipulated, and it was
observed that the characteristic size of the crystals, and their time
of formation decrease as the size of the unit cell increases.
Using the proposed exponential dependences of crystal size and time of
formation upon the free energy barrier, the second author,
together with M. Grabe, G. Oster, and P. Nollert \cite{GNON03}
quantified the decrease of free energy barrier with increasing size of
unit cell. 
The results are consistent with an independent estimate of the free
energy barrier based upon the elastic energy of embedding proteins in the
bilayer.

Certain detailed results of the current paper are expected to be less
robust, and should be regarded as documenting the predictions of
(oversimplified) classical kinetics.
In particular, the emerging distribution of cluster sizes after
nucleation becomes conspicuously narrow, so the sizes of clusters are
much closer to one uniform size than is observed in experiments.
One proposal \cite{WOLFER03:personal} is that a broader distribution results if in fact we
observe the superposition of distributions in a spatially
inhomogeneous nucleation process. 

The sharp front of the cluster size distribution at the largest
cluster size as predicted by our model is also expected to be a
casualty of any comparison with reality.
Many effects could regularize it.
In particular, B. Niethammer and J.J.L. Velasquez \cite{NV06} formulated a
diffusion-like correction to the advection PDE of the LS model, based
on screening fluctuations in the local super-saturation seen about
clusters. 

The extreme sensitivity of the nucleation rate upon small changes
in the chemical potential \mbox{$\eta$} of monomers was exploited in our asymptotic
solution of the creation transient.  
But this extreme sensitivity is a potential source of difficulties as
well.  
For instance, most works to date, such as Penrose \cite{PEN97}, Niethammer and
Velasquez \cite{NV06}, use the approximation to the conservation of particles,
which says:  ``Density of monomers plus density of `large' clusters equals
total particle density.'' 
In our work, this simple approximation to particle conservation is
retained, so as to not distract from the main results.
However, even small corrections to the conservation law and
a consequential small change in \mbox{$\eta$} can be amplified to
large (relative) corrections to the nucleation rate, which is
exponentially small as \mbox{$\eta$} goes to zero.  
Specifically, Wu in \cite{Wu96} mentions the quasi-equilibrium distribution of
sub-critical clusters (`embryos' in his terminology) with \mbox{$1< n < n_*$}.
We propose that the inclusion of the embryos in the conservation of
particles is one of those small corrections that lead to
significant changes in predictions of the nucleation rate.

\section{Acknowledgments}
This paper was supported in part by the National Science Foundation.
Y. Farjoun was partially supported by grants DMS--0515616 and DMS--0703937.
The authors would like to thank the anonymous referees for their helpful
suggestions, and specifically for the title of the paper.
\section*{Appendix}
\appendix
\renewcommand{\thesection}{\Alph{section}}
\renewcommand{\theequation}{\Alph{section}.\arabic{equation}}

\section{Numerical Solution of Integral Equation}
\label{sec:numerical:explain}
\begin{figure}[ht!]
\begin{center}
\resizebox{5.5cm}{!}{\includegraphics{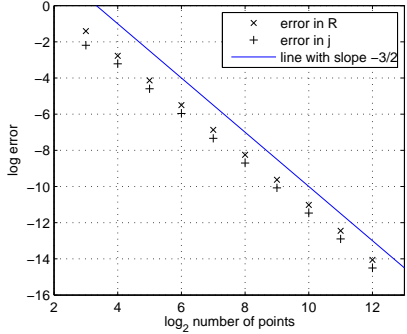}}
\caption[Nucleation rate as a function of time]{The errors in $j$ and
  $R$, together with a line of slope $-3/2$ in a log-log plot. The
  x-axis is the log (base 2) of the number of points used, and the
  y-axis is the (natural) log of the error.} 
\label{fig:errors}
\end{center}
\end{figure}

The analysis of the nucleation era requires the solution of the
integral equation,
\begin{equation}
  \delta\eta(t)=-\int\limits_0^t \brk{\frac23\brk{t-\tau}}^\frac32
  \E^{\delta\eta(t)}\ud{\tau}.\label{eq:integral:equa}
\end{equation}
We find anapproximate solution of this equation on a set
of equally spaced times $t_n$ on the interval $[0,\, 15]$. 
The value of \mbox{$\delta\eta(t_{n+1})$}  is calculated from the integral of
\mbox{$\delta\eta$} up to time \mbox{$t_{n+1}$} using the trapezoidal rule approximation. 
This is not implicit as the \mbox{$t-\tau$} term in the
integrand of \eqref{eq:integral:equa} vanishes at \mbox{$\tau=t$}. 
The order of accuracy of the method is found to be \mbox{$\frac32$}
(see Fig.~\ref{fig:errors}). 
This fractional order is probably due to the cusp in the integrand
at \mbox{$t=\tau$}, which adds an error of  \mbox{$\O(\Delta t)^{3/2}$} to the integral. 
The convergence of the error can be seen in Fig.~\ref{fig:errors}
where the (natural) log of the errors (of both $j$ and $R$) are plotted
against the log (base 2) of the number of points in the segment  \mbox{$[0,\,15]$}.
The most accurate result we have obtained---with $2^{13}$ points---for
the resulting density of clusters is $R=1.7109162\pm3$. 
\section{Validity of the quasi-steady-state assumption}
\label{app:qss}
One of the assumptions used in this paper is that the nucleation rate corresponds to the Zeldovich rate with the instantaneous super-saturation. 
This ``quasi-steady-state'' (QSS) assumption has been previously studied by Shneidman \cite{Shneidman87}. It was shown that the QSS assumption is valid if
\begin{align}
\brk{\frac{G_*}{k_BT}}^{-\gamma}&\Approx 1
\label{eq:qss:factor}
\intertext{where $\gamma$ is given by}
\gamma=-\tau\frac{d}{dt}\brk{\frac{G_*}{k_BT}}&,\qquad \tau = \left.\frac{d\dot n}{dn}\right|_{n=n_*}.\label{eq:qss:factor:2}
\end{align}
In the case of the creation era, we have shown that the characteristic time $[t]$ is exponentially long, as given by \eqref{eq:scale:t}. 
Therefore the derivative of the free-energy barrier is exponentially small. 
In \eqref{eq:qss:factor:2}, $\tau$ is a parameter that depends on the physical model for the growth of small clusters. 
For example, the Becker-D\"oring model, has $\tau=\frac13\frac{\omega\eta}{\sig}$.
Thus, we see that the exponent $\gamma$ is itself exponentially close
to zero (due to the time derivative), and the LHS in \eqref{eq:qss:factor} is extremely close to~1. 
This validates, a-posteriori, the QSS assumption.

\section{Critical size $n_*$ is not equal to zero}
\label{app:n_*}
The effective boundary condition \eqref{eq:advection:BC} results from a heuristic \emph{asymptotic matching} in the small super-saturation limit $\eta\goto0$: 
The Zeldovich rate on the RHS is a quasi-static discrete flux which measures the net rate of creation of $n+1-$clusters from $n-$clusters per unit volume, valid for $n$ on the order of critical size $n_{*}=\frac{\sig^{3}}{\eta^{3}}$.
The LHS is a continuum approximation to the flux based on diffusion limited growth.
Presumably, diffusion limited growth is valid for some range of ``large'' cluster sizes with $n\gg n_{*}$. 
Of course we assume that the characteristic size $[n]\propto \E^{\frac35\frac{\sig^{3}}{2\eta^{2}}}\ll n_{*}$ is in this range.
Moreover, we assume that diffusion limited growth holds asymptotically for clusters of an ``intermediate'' size $n_{+}=n_{+}(\eta)$ so $n_{*}\ll n_{+}\ll[n]$.
Consider clusters with size $n$ in the interval $n_{*}<n<n_{+}$. 
Under quasi-static conditions, we expect that influx through the $n=n_{*}$ end at the Zeldovich rate balances the ``diffusive limited'' outflux through the $n=n_{+}$ end, so that
\begin{equation}
\Omega \E^{-\frac{\sig^{3}}{2\eta^{2}}}=d\eta n_{+}^{\third}r(n_{+},t).
\label{eq:BC:n_+}
\end{equation}
Since $n_{+}/[n]\goto 0$ as $\eta\goto0$, we obtain \eqref{eq:advection:BC} as the formal limit of \eqref{eq:BC:n_+} as $\eta\goto0$.

The appearance of zero as the lower limit of the integral $\int_{0}^{\infty}nr(n,\,t)\ud{n}$ in (2.6) is also part of the small super-saturation limit $\eta\goto 0$.
The exact density of monomers in clusters is the discrete sum $\sum_{n=2}^{\infty}n f_{n}$, where $f_{n}$ is the density of $n-$clusters. 
The standard idea behind replacing this sum with an integral is that the $f_{n}$ are the values of a smooth function at integer arguments, whose characteristic scale of the independent variable is much larger than unity.
That is \emph{almost} what we have.
We approximate $f_{n}$ by $r(n,\,t)$ whose characteristic scale of $n$ is $[n]\propto\E^{\frac35\frac{\sig^{3}}{2\eta^{2}}}\goto\infty$ as $\eta\goto0$.
Of course the continuum approximation to $f_{n}$ breaks down for some range of $n$ with $n\ll[n]$, so in addition we are \emph{assuming} that the contributions to the sum $\sum_{n=0}^{\infty}nf_{n}$ from this $n\ll[n]$ range are negligible as $\eta\goto0$.

\bibliographystyle{amsplain}
\bibliography{general}
\listoffigures
\end{document}